\RequirePackage{fix-cm}
\documentclass[smallextended]{svjour3}       
\smartqed  
\usepackage{cite}
\usepackage{natbib}
\usepackage{url}
\usepackage{amsmath,amssymb,amsfonts}
\usepackage{algorithmic}
\usepackage{graphicx}
\usepackage{textcomp}
\usepackage{xcolor}
\usepackage{balance}
\usepackage{booktabs}
\usepackage{pgfplots}
\usepackage{subfig}
\usepackage{tabularx}
\usepackage{tikz}
\usepackage{caption}
\usepackage{alltt}
\usepackage{comment}
\usepackage[T1]{fontenc}

\definecolor{author}{rgb}{1.0, 0.6, 1.0}
\definecolor{year}{rgb}{0.8, 0.6, 0.6}
\definecolor{title}{rgb}{0.6, 0.6, 1.0}
\definecolor{booktitle}{rgb}{0.6, 0.8, 0.6}
\definecolor{volume}{rgb}{0.4, 1.0, 0.8}
\definecolor{number}{rgb}{0.8, 0.6, 1.0}
\definecolor{pages}{rgb}{1.0, 0.6, 0.6}
\definecolor{url}{rgb}{0.6, 1.0, 0.8}

\newcommand\hlauthor[2][author]{\hspace*{-\fboxsep}\colorbox{#1}{#2}\hspace*{-\fboxsep}}
\newcommand\hlyear[2][year]{\hspace*{-\fboxsep}\colorbox{#1}{#2}\hspace*{-\fboxsep}}
\newcommand\hltitle[2][title]{\hspace*{-\fboxsep}\colorbox{#1}{#2}\hspace*{-\fboxsep}}
\newcommand\hlbook[2][booktitle]{\hspace*{-\fboxsep}\colorbox{#1}{#2}\hspace*{-\fboxsep}}
\newcommand\hlvolume[2][volume]{\hspace*{-\fboxsep}\colorbox{#1}{#2}\hspace*{-\fboxsep}}

\newcommand\hlurl[2][url]{\hspace*{-\fboxsep}\colorbox{#1}{#2}\hspace*{-\fboxsep}}

\def\BibTeX{{\rm B\kern-.05em{\sc i\kern-.025em b}\kern-.08em
    T\kern-.1667em\lower.7ex\hbox{E}\kern-.125emX}}
\begin{document}
\begin{sloppy}
\title{From Academia to Software Development: Publication Citations in Source Code Comments}


\author{Akira Inokuchi \and Yusuf Sulistyo Nugroho \and Supatsara Wattanakriengkrai \and Fumiaki Konishi \and Hideaki Hata \and Christoph Treude \and Akito Monden \and Kenichi Matsumoto
}


\institute{Akira Inokuchi, Yusuf Sulistyo Nugroho, Supatsara Wattanakriengkrai, Fumiaki Konishi, Hideaki Hata, Kenichi Matsumoto \at
              Nara Institute of Science and Technology \\
              \email{\{inokuchi.akira.hw0, yusuf\_sulistyo.nugroho.yi5, wattanakri.supatsara.ws3,
              konishi.fumiaki.jw8, hata, matumoto\}@is.naist.jp}            
           \and
           Christoph Treude \at
              University of Adelaide    \\
              \email{christoph.treude@adelaide.edu.au}
           \and
           Akito Monden \at
              Okayama University    \\
              \email{monden@okayama-u.ac.jp}
}

\date{Received: date / Accepted: date}

\maketitle

\begin{abstract}
Academic publications have been evaluated in terms of their impact on research communities based on many metrics, such as the number of citations. On the other hand, the impact of academic publications on industry has been rarely studied. This paper investigates how academic publications contribute to software development by analyzing publication citations in source code comments in open source software repositories.
We propose an automated approach for detecting academic publications based on Named Entity Recognition, and achieve 0.90 in $F_1$ as detection accuracy.
We conduct a large-scale study of publication citations with 319,438,977 comments collected from 25,925 active repositories written in seven programming languages.
Our findings indicate that academic publications can be knowledge sources for software development.
These referenced publications are particularly from journals.
In terms of knowledge transfer, algorithm is the most prevalent type of knowledge transferred from the publications, with proposed formulas or equations typically implemented in methods or functions in source code files.
In a closer look at GitHub repositories referencing academic publications, we find that science-related repositories are the most frequent among GitHub repositories with publication citations, and that the vast majority of these publications are referenced by repository owners who are different from the publication authors.
We also find that referencing older publications can lead to potential issues related to obsolete knowledge.
\keywords{publication citations \and source code comments \and open science}
\end{abstract}

\section{Introduction}
\label{sec:intro}
How has ``literate programming'', a programming paradigm proposed by~\citet{Knuth:1984:LP:473.479},
impacted software development?
Although the paper~\citep{Knuth:1984:LP:473.479} has been referenced by more than 2,200 academic papers\footnote{2,289 citations. Google Scholar: accessed Apr. 2020.}, can we see its impact
on real-world programming?
Similar to publication citations in academic papers, we observe that developers also reference academic publications when implementing code.
For example, when we searched for the keyword ``literate programming'' in a source code search engine \texttt{searchcode\footnote{searchcode: \url{https://searchcode.com/}, accessed Apr. 2020.}},
we found a comment in a Java source file\footnote{openesb-components/ojc-core/component-common/xmlbeans/xbean-src/2.3.0/xmlbeans/test/tools/src/tools/util/Diff.java} as shown in Figure~\ref{fig:example}.
From Vol. 30, No. 7 in 1987 to Vol. 33, No. 3 in 1990, columns on ``Literate Programming'' were published in the Communications of
ACM, and the above-mentioned code comment was written based on one such article~\citep{VanWyk:1989:LP:63526.315960}.

\begin{figure*}
\begin{verbatim}
     * This program is intended to be pedagogic.  Specifically, this program was
     * the basis of the Literate Programming column which appeared in the
     * Communications of the ACM (CACM), in the June 1989 issue (32, 6,
     * 740-755).
 \end{verbatim}
\caption{A code comment referencing the paper~\citep{VanWyk:1989:LP:63526.315960}.}
\label{fig:example}
\end{figure*}

In this work, we study such publication citations in source code comments.
In academia, publication citations can be considered as a function in scientific communication among texts, as an indicator of reward in the science system, and as the collective character of scientific achievements~\citep{Leydesdorff1998}.
Since software developers are likely to reference external knowledge sources to implement code~\citep{Hata2019}, publication citations like the above example can be considered as knowledge transfer from academia to practice. 
The significance of this work is related to software documentation~\citep{8094446} from the perspective of knowledge sharing~\citep{7000568,8359344}.
For improving documentation and mitigating potential issues of obsolete knowledge, understanding the nature of publication citations in software development is necessary and important.

This work is related to research on source code comments in terms of software documentation.
Source code comments have been found to document personal and team tasks~\citep{Storey:2008:TBE:1368088.1368123} and technical debt~\citep{potdar2014exploratory}, among others.
A recent study of links in source code comments revealed that comments are used to express background meta information, source code context information, and technical debt, referring to external sources via links~\citep{Hata2019}.
Similar to links, referencing academic publications indicates external sources to explain associated code. 
However, to the best of our knowledge, publication citations in source code comments have not been studied comprehensively so far, hence there is a lack of knowledge on what kinds of academic achievements are referenced when developing software.
To address this knowledge gap, we conducted an empirical study to explore publication citations in source code comments.

Considering the detection of publication citations in source code comments as an information extraction task, there are three challenges.
First, there are no explicit keywords to search. While regular expressions such as \texttt{/http$\backslash$S+/} can be used to identify links in text~\citep{Hata2019}, we cannot prepare such regular expressions to identify publication citations because of nonexistence of common words and patterns.
Second, references to publications in source code comments are written in free format and can appear anywhere. 
Since there is no rule for publication citations in source code comments, developers freely express references, such as omitting titles and other entities or using \BibTeX~style, which makes it difficult to apply heuristics in parsing citations~\citep{Giles:1998:CAC:276675.276685}.
The third challenge is the rarity of publication citations.
In research on self-admitted technical debt, several machine learning approaches have been proposed to detect comments indicating self-admitted technical debt~\citep{7820211,Huang:2018:IST:3188697.3188709,8661216}. These approaches are based on supervised classification utilizing large amounts of labeled data. Since there are not many comments referencing publications (we estimate less than $0.1\%$ of all comments in our data), it is difficult to prepare enough data for training.
Our initial trial for automating detection with supervised classification, i.e., classifying comments in terms of whether they contain citations, did not work well due to these difficulties.
We also tried preliminary manual investigation by using public code search services. 
However, manual investigation is not scalable and the results can be biased.

To conduct a large-scale empirical study and considering these challenges related to information extraction, we developed an approach for detecting publication citations in source code comments based on Named Entity Recognition (NER)~\citep{C18-1182}.
By identifying publication-related named entities, such as authors, titles, journal names, years, etc., we infer the existence of publication citations. A model trained on manually annotated publication citations in comments was used to detect publication citations from 319,438,977 distinct comments in active software development projects written in seven programming languages.
Based on manual validation with a statistically representative sample from all the comments we detected, we obtained 0.90 in $F_1$ for detecting publication citations.
A manual investigation of 296 identified comments in our sample shows that academic papers published in journals are more frequently referenced in source code comments than those published in other types of venues.
The main objective of referencing these academic publications is to reference knowledge sources for source code implementation.
Algorithm is the most common knowledge type transferred from the referenced publications, with proposed formulas or equations typically implemented in methods or functions in source files.
In term of software domains, science-related repositories are the most frequent among GitHub repositories with publication citations.
The vast majority of these publications are referenced by repository owners other than the publication authors.
We sometimes find multiple publications referenced in a single comment and that \textsf{ACM Transactions on Mathematical Software} and \textsf{Communications of the ACM} are the most referenced journals, with publications from the 1990s and 2000s most frequently referenced. We also identify potential issues related to referencing older publications for potentially obsolete knowledge.

In summary, the contributions of this paper are three-fold:
\begin{itemize}
    \item an approach for the automated detection of publication citations in source code comments based on Named Entity Recognition,
    \item a large-scale and comprehensive study of publication citations in active software development projects written in seven programming languages (C, C++, Java, JavaScript, Python, PHP, and Ruby), and
    \item qualitative and quantitative analyses of detected publication citations to understand knowledge transfer from academia to software development. Our results are publicly available (Section \ref{ssec:appendix})
\end{itemize}

\section{Preliminary Manual Investigation}
\label{sec:manual}

\begin{table}
\centering
\caption{The number of referenced papers per journal in C and Java source code comments}
\label{tab:journal}
\begin{tabular}{lrr}
\toprule
\textbf{journal} & \textbf{C} & \textbf{Java} \\
\midrule
COMMUN ACM & 78 & 52 \\
ACM T MATH SOFTWARE & 78 & 23 \\
IEEE COMPUT GRAPH & 16 & 1 \\
ACM T GRAPHIC & 9 & 11 \\
ACM T PROGR LANG SYS & 11 & 4 \\
J ACM & 6 & 7 \\
ACM COMPUT SURV & 2 & 6 \\
SIAM J COMPUT & 1 & 6 \\
IEEE T SOFTWARE ENG & 2 & 5 \\
IEEE T SYST MAN CY A & 3 & 4 \\
ACM T COMPUT SYST & 2 & 2 \\
\bottomrule
\end{tabular}
\end{table}

\begin{table*}
    \centering
    \caption{Frequently referenced papers in Communications of the ACM in decades}
    \label{tab:decade}
    \resizebox{\columnwidth}{!}{%
    \begin{tabular}{lrp{25pc}rp{7pc}}
        \toprule
        \textbf{decade} & \textbf{rank} & \textbf{paper} & \multicolumn{2}{l}{\textbf{\# code citations}} \\
        \midrule
        1950s & 1 & D. L. Shell, \textbf{A high-speed sorting procedure} (1959) & 2 & (C 1, Java 1) \\
        \midrule
        1960s & 1 & Robert L. Smith, \textbf{Algorithm 116: Complex division} (1962) &  21 & (C 19, Python 2) \\
        & 2 & Immo O. Kerner, \textbf{Algorithm 283: Simultaneous displacement of polynomial roots if real and simple} (1966) & 15 & (C 14, Java 1) \\
        & 3 & Robert G. Tantzen, \textbf{Algorithm 199: conversions between calendar date and Julian day number} (1963) & 8 & (C 6, C++ 2) \\
        & 4 & G. Marsaglia, \textbf{Generating discrete random variables in a computer} (1963) & 6 & (C 3, Java 3) \\
        & 5 & Paul Friedland, \textbf{Algorithm 312: Absolute value and square root of a complex number} (1967) & 5 & (C 5) \\
        \midrule
        1970s & 1 & Jay Earley, \textbf{An efficient context-free parsing algorithm} (1970) & 29 & (Python 29) \\
        & 2 & J. P. Chandler and W. C. Harrison, \textbf{Remark on algorithm 201 [M1]: shellsort} (1970) & 11 & (C 11) \\
        & 3 & J. H. Ahrens and U. Dieter, \textbf{Computer methods for sampling from the exponential and normal distributions} (1972) & 8 & (C 6, C++ 2) \\
        & 3 & Alfred V. Aho and Margaret J. Corasick, \textbf{Efficient string matching: an aid to bibliographic search} (1975) & 8 & (C 8) \\
        & 4 & Jack Bresenham, \textbf{A linear algorithm for incremental digital display of circular arcs} (1977) & 6 & (C 6) \\
        & 4 & R. C. H. Cheng, \textbf{Generating beta variates with nonintegral shape parameters} (1978) & 6 & (C 5, C++ 1) \\
        & 5 & Michael A. Malcolm, \textbf{Algorithms to reveal properties of floating-point arithmetic} (1972) & 5 & (C 3, C++ 1, Java 1) \\
        & 5 & W. Morven Gentleman and Scott B. Marovich, \textbf{More on algorithms that reveal properties of floating point arithmetic units} (1974) & 5 & (C 3, C++ 1, Java 1) \\
        & 5 & W. R. Franta and Kurt Maly, \textbf{An efficient data structure for the simulation event set} (1977) & 5 & (C 5) \\
        \midrule
        1980s & 1 & E. R. Fiala and D. H. Greene, \textbf{Data compression with finite windows} (1989) & 72 & (C 72) \\
        & 2 & S. K. Park and K. W. Miller, \textbf{Random number generators: good ones are hard to find} (1988) & 42 & (C 25, C++ 13, C\# 2, Java 1, Python 1) \\
        & 3 & Reinhold P. Weicker, \textbf{Dhrystone: a synthetic systems programming benchmark} (1984) & 36 & (Python 30, C 4, C++ 2) \\
        & 4 & Per-Ake Larson, \textbf{Dynamic hash tables} (1988) & 13 & (C 11, C++ 1, Java 1) \\
        & 5 & Richard J. Cichelli, \textbf{Minimal perfect hash functions made simple} (1980) & 8 & (C 8) \\
        & 5 & J. H. Ahrens and U. Dieter, \textbf{Generating gamma variates by a modified rejection technique} (1982) & 8 & (C 4, C++ 2, Java 2) \\
        & 5 & Ian H. Witten, Radford M. Neal, and John G. Cleary, \textbf{Arithmetic coding for data compression} (1987) & 8 & (C 7, C++ 1) \\
        \midrule
        1990s & 1 & Bala R. Vatti, \textbf{A generic solution to polygon clipping} (1992) & 15	& (C++ 11, C\# 4) \\
        & 2 & Peter K. Pearson, \textbf{Fast hashing of variable-length text strings} (1990) & 10 & (C 7, C++ 3) \\
        & 3 & William Pugh, \textbf{Skip lists: a probabilistic alternative to balanced trees} (1990) & 8 & (C 7, Python 1) \\
        & 4 & David F. Carta, \textbf{Two fast implementations of the ``minimal standard'' random number generator} (1990) & 7 & (C 2, C++ 5) \\
        & 5 & Edward A. Fox, Lenwood S. Heath, Qi Fan Chen, and Amjad M. Daoud, \textbf{Practical minimal perfect hash functions for large databases} (1992) & 4 & (C 3, Java 1) \\
        \midrule
        2000- & 1 & George Marsaglia, \textbf{Seeds for random number generators} (2003) & 1 & (C++ 1) \\
        \bottomrule
    \end{tabular}%
    }
\end{table*}

We started manual investigation in 2013 using a public code search service for OSS, \texttt{Ohloh}, provided by Black Duck Software at that time\footnote{Ohloh was discontinued in 2016.}.
It covered around 80 thousand C projects and 70 thousand Java projects.
Using the code search service, the fourth author searched 50 journal names in C and Java source code, then manually validated publication citations. Targeted journals were selected as top 50 aggregate impact factors from 2004 to 2012 in the categories of \textsf{COMPUTER SCIENCE, SOFTWARE} and \textsf{COMPUTER SCIENCE, THEORY \& METHODS} in Journal Citation Reports. Several abbreviations of journal names were prepared as search queries.
Table~\ref{tab:journal} summarizes the number of distinct papers found in C and Java source code comments.
\textsf{COMMUNICATIONS of the ACM} and \textsf{ACM TRANSACTIONS on MATHEMATICAL SOFTWARE} were found to be most referenced journals.
Regarding \textsf{IEEE Transactions on Software Engineering}, some examples are available on GitHub\footnote{\url{https://github.com/svn2github/cytoscape/blob/a3df8f63dba4ec49942027c91ecac6efa920c195/csplugins/trunk/agilent/kuchinsky/infovis_0.9beta/src/infovis/tree/visualization/nodelink/RTLayout.java#L25}}~\footnote{\url{https://github.com/sba1/ontologizer/blob/62765f4e5d42ddaae999c480e7ef50bb510f5fc8/ontologizer.grappa/src/main/java/att/grappa/Grappa.java#L13}}.

In 2016, the fifth author again manually investigated publication citations of \textsf{COMMUNICATIONS of the ACM}, using the public code search service \texttt{searchcode} with the query ``cacm''.
Data sources of \texttt{searchcode} include Bitbucket, GitHub, Google Code, Sourceforge, and GitLab, among others, covering various programming languages.
Table~\ref{tab:decade} shows frequently referenced papers across different decades.
In this manual investigation, five programming languages were targeted (C, C++, C\#, Java, and Python) to see how papers are referenced in different languages.
We found that although the majority of referencing comments appeared in C source code, developers using other languages also referenced academic publications. In addition, various papers had been utilized to implement code from the 1950s to the 2000s, which may indicate a risk of obsolete knowledge and code, especially for source code files referencing older publications.

From our manual investigation, we see the existence of publication citations in source code comments.
However, such manual investigation does not scale; we cannot conduct large-scale empirical studies.
Using code search services also has limitations, since these services run on predetermined data sources, preventing us from controlling criteria for selecting software to be analyzed. 
In addition, choosing search queries can bias results.
To address these issues, we developed an automated approach for detecting academic publications and conducted a large-scale study of publication citations in source code comments.

\section{Study Setup}

We consider the detection of publication citations in source code comments as a Named Entity Recognition (NER) task~\citep{C18-1182}.
We try to detect publication citations by identifying sets of publication-related named entities, such as authors, titles, journal names, and years.
For our empirical study, we collect a large amount of source code comments.

In the following, we describe our study approach, data collection procedure, and named entity recognition.

\subsection{Overview}

\begin{figure*}
\centering
\includegraphics[width=\linewidth]{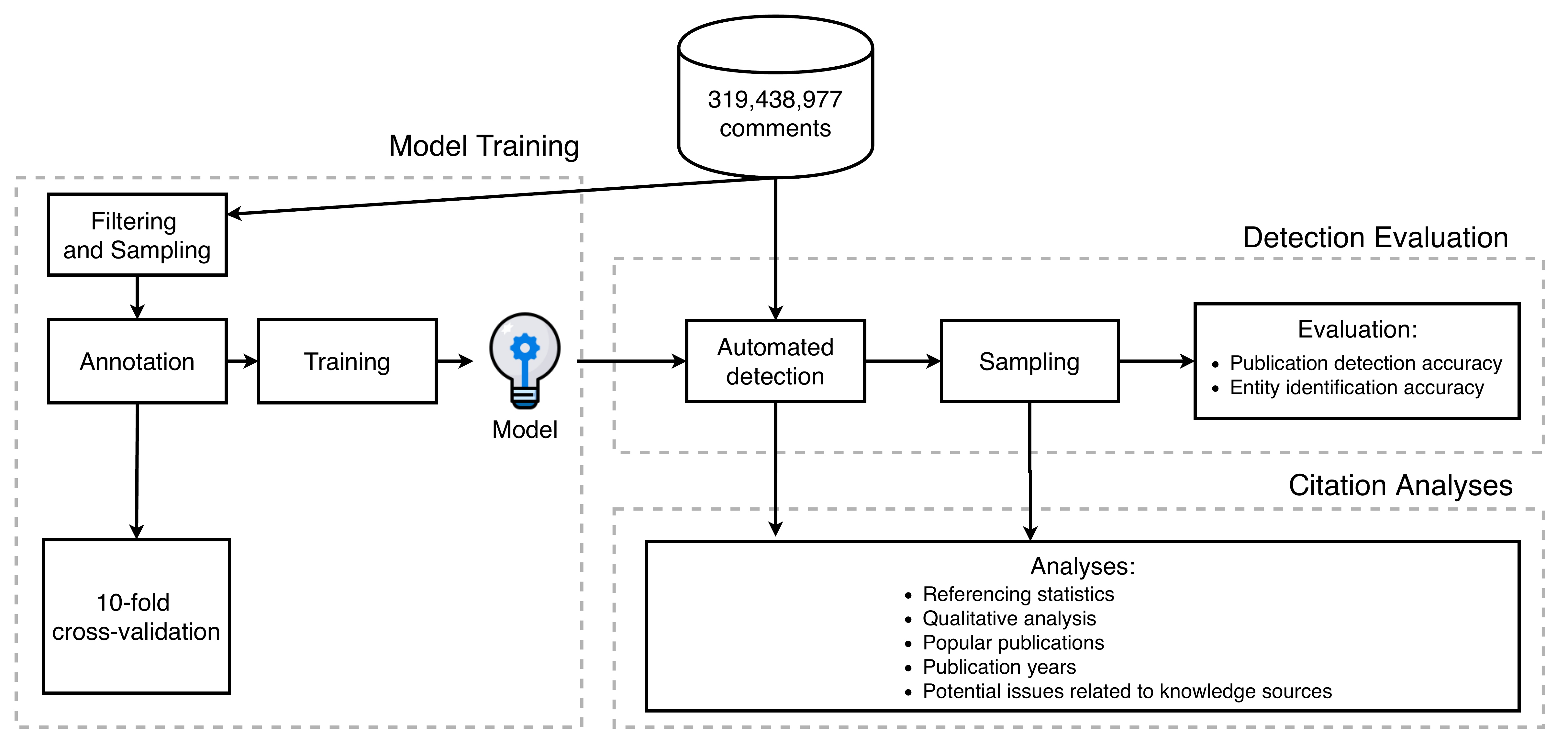}
\caption{An overview of our large-scale study of publication citations in source code comments from detection to analysis.}
\label{fig:extractionprocess}
\end{figure*}

Figure~\ref{fig:extractionprocess} presents an overview of our large-scale study of publication citations in source code comments, which consists of three components:
training an NER model with manually annotated citations in specific comments (Section~\ref{sec:train}),
evaluation of publication citation detection in a large amount of comments (Section~\ref{sec:evaluation}), and
analysis of the detected citations (Section~\ref{sec:analysis}).

\subsection{Data Collection}

\begin{table}
\centering
\caption{Collected distinct comments}
\label{tab:comments}
\begin{tabular}{lrr}
\toprule
& \textbf{\# repositories} & \textbf{\# comments} \\
\midrule
C & 2,482 & 217,117,412 \\
C++ & 3,211 & 39,266,272 \\
Java & 4,472 & 22,638,116 \\
JavaScript & 6,224 & 15,056,958 \\
Python & 4,715 & 8,059,852 \\
PHP & 2,827 & 16,085,456 \\
Ruby & 1,994 & 1,214,911 \\
\midrule
\textbf{sum} & \textbf{25,925} & \textbf{319,438,977} \\
\bottomrule
\end{tabular}
\end{table}

Source code comments were collected following the same procedure as in previous work~\citep{Hata2019}, which targeted active software development repositories on GitHub written in seven common programming languages, that is, C, C++, Java, JavaScript, Python, PHP, and Ruby. 
These seven languages are considered to be common since they have been consistently ranked in the top 10 languages on GitHub in the recent 10 years~\citep{github2015blog,githut,github2017octoverse}.

Active software development repositories were selected from the GHTorrent dataset\footnote{MySQL database dump 2018-04-01 from \url{http://ghtorrent.org/downloads.html}.}~\citep{Gousios:2013:GDT:2487085.2487132} with the following criteria~\citep{Hata2019}: (i) more than 500 commits (the same threshold used in previous work~\citep{Aniche:2018:CSM:3238579.3238606}), and (ii) at least 100 commits in the most active two years (to remove long-term less active projects and short-term repositories, which may not be software development projects~\citep{Munaiah:2017:CGE:3147777.3147808}).

Table~\ref{tab:comments} shows the total number of collected repositories per language and in total. 
From the collected repositories, code comments were extracted in the \texttt{HEAD} commits with the tool~\citep{hideaki_hata_2019_2550683}\footnote{Comment Lister. \url{https://github.com/takashi-ishio/CommentLister}} used in previous work~\citep{Hata2019}. 
Since there are many duplicate comments mainly because of code reuse, we obtained distinct code comments for each language. 
As seen in Table~\ref{tab:comments}, we obtained more than 300 million distinct code comments, with the largest amount of comments associated with C code. 
From these comments, special characters (`\textbackslash n', `/', `*', `\textbackslash', `\#', `!') were removed as part of a pre-processing step.

Compared to using \texttt{searchcode}, our dataset consists of source code comments without duplicates only from active repositories, but is limited to seven programming languages.

\subsection{Named Entity Recognition}

Named Entity Recognition (NER) is the task of identifying named entities in unstructured text, such as the name of a person, location, organization, or time~\citep{C18-1182}.
In the fields of Natural Language Processing (NLP) and Information Retrieval (IR), NER is used to help in answering real-world questions, summarizing text, and translating.
We adopt \texttt{Spacy}~\citep{D15-1162} as a NER tool.
Spacy is a module written in Python and Cython for NLP to train NER models\footnote{\url{https://spacy.io}}.
It provides features that are commonly used in NLP projects to identify named entities, such as tokenization, lemmatisation, part-of-speech tagging, and entity recognition.
One benefit of using Spacy is having better performance compared to NLTK, especially on tokenization and part-of-speech tagging tasks when applied to software-related texts~\citep{Omran:2017:CNL:3104188.3104213}.

When using Spacy for NER, we can make use of publicly available pre-trained statistical models\footnote{\url{https://spacy.io/models/en}}. 
We selected \texttt{en\_core\_web\_sm}, a model trained on \texttt{OntoNotes 5}. 
OntoNotes 5 is the final release of the OntoNotes project, which intends to annotate a large corpus comprising various genres of text (news, conversational telephone speech, weblogs, usenet newsgroups, broadcast, talk shows)~\citep{ontonote5}. 
Several entity types are supported in the model, such as person, organization, and date\footnote{\url{https://spacy.io/api/annotation#named-entities}}.

\subsection{Online Appendix}
\label{ssec:appendix}

We make our datasets and results publicly available.
Our online appendix contains
manually annotated publication-related entities (Section~\ref{sec:annotation}),
our manual identification of publication citations in a sample (Section~\ref{sec:sample}), and
our qualitative analysis results (Section~\ref{sec:qa}).
The appendix is available at \url{https://github.com/supatsara-wat/PaperCitationCodeComments}.

\section{Model Training}
\label{sec:train}

To apply NER for identifying publication-related named entities, we prepare new entity types and annotate a set of comments with them. These annotated comments are used to train a model.

\subsection{Filtering and Sampling}

\begin{table}
    \centering
    \caption{Comments including keywords ACM and IEEE}
    \label{tab:data}
    \begin{tabular}{llrrr}
    \toprule
     & & \textbf{ACM} & \textbf{IEEE} & \textbf{sum} \\
     & & 5,521 & 16,682 & 22,203 \\
    \midrule
    group A & all & 4,372 & 7,656 & 12,028 \\
     & sample & 353 & 366 & 719 \\
     & cite & 175 & 92 & 267 \\
     & none & 178 & 274 & 452 \\
    \midrule
    group B & all & 1,149 & 9,026 & 10,175 \\
     & sample & 288 & 369 & 657 \\
     & cite & 1 & 4 & 5 \\
     & none & 287 & 365 & 652 \\
    \midrule
    train \& validate & sample sum & 641 & 735 & 1,376 \\
     & cite & 176 & 96 & 272 \\
     & none & 465 & 639 & 1,104 \\
    \bottomrule
    \end{tabular}
\end{table}

For preparing data to be used to train our model, comments that contain publication citations are needed.
From the entire set of 319,438,977 comments, we first collect comments that include popular association names, ACM and IEEE, since there can be various journals or conferences related to these associations referenced in source code comments as seen in Section~\ref{sec:manual}. As shown in Table~\ref{tab:data}, we obtained 5,521 distinct comments with the keyword `acm' and 16,682 distinct comments with the keyword `ieee'.

We divide these comments into two groups A and B, as a group of comments that are prone to contain publication citations and a group of comments that are not prone to contain them.
The first author manually investigated these comments and designed the criteria for this grouping.
For comments containing `acm', group B comments were identified as containing the keyword ``@acm.org'' as they tend to contain only such email addresses. The other 4,372 comments were assigned to group A.
Similarly, ``IEEE.org'' and ``IEEE STD'' were used as keywords to select comments containing `ieee' for group B.
To identify comments referring to IEEE Standards and putting them into group B, we considered numbers \textit{488}, \textit{754}, \textit{802}, \textit{854}, \textit{1003}, \textit{1076}, \textit{1149}, \textit{1275}, \textit{1284}, \textit{1355}, and \textit{1363}, following \textit{IEEE` '} (space), \textit{IEEE-} or \textit{IEEE\_}.
For comments containing `ieee', there are 9,026 comments in group B and 7,656 comments in group A.

For these four groups (two groups each for ACM and IEEE), we obtained a statistically representative sample from each group.
The required sample size was calculated so that conclusions related to the ratio of publication citations would generalize to all comments in the same group with a confidence level of 95\% and a confidence interval of 5\footnote{\url{https://www.surveysystem.com/sscalc.htm}}.

\subsection{Annotation}
\label{sec:annotation}

The first author manually investigated all 1,376 comments (353+366+288+369) and identified 272 comments with publication citations (175+92+1+4). 
For these 272 comments including publication citations, we annotated the following publication-related entities (each number presents the number of appearance in all citations in the comments):
\textbf{author} (888),
\textbf{title} (408),
\textbf{year} (407),
\textbf{booktitle\_or\_journal} (353),
\textbf{pages} (251),
\textbf{volume} (234),
\textbf{number} (165),
\textbf{month} (116),
\textbf{url} (92),
\textbf{publisher} (74),
\textbf{address} (31),
\textbf{doi} (27),
\textbf{isbn} (7), and
\textbf{issn} (2).
Since individual authors are annotated separately, the number of \textit{author} entities is the largest.
Book titles and names of journals or conferences are integrated into one entity \textit{booktitle\_or\_journal}.
For \textit{volume}, \textit{number}, and \textit{pages}, some citations contain keywords, such as `Vol.', `No.', and `pages'. In such cases, these keywords were included for entities as well as numbers.

\subsection{Validation and Training}

\begin{table}
\centering
\caption{Accuracy with the combination of detected entities}
\label{tab:combination}
\begin{tabular}{lrrr}
\toprule
\textbf{set of entities} & \textbf{precision} & \textbf{recall} & \textbf{$F_1$} \\
\midrule
author, title, year, booktitle\_or\_journal & \textbf{0.99} & 0.78 & 0.87 \\
title, year, booktitle\_or\_journal & 0.96 & 0.81 & 0.88 \\
author, year, booktitle\_or\_journal & 0.96 & 0.80 & 0.87 \\
author, title, booktitle\_or\_journal & 0.95 & 0.83 & 0.89 \\
year, booktitle\_or\_journal & 0.92 & 0.84 & 0.88 \\
\bottomrule
\end{tabular}
\end{table}

To evaluate the effectiveness of NER with our annotated comments for publication citation detection, we conduct 10-fold cross-validation with all 1,376 comments shown in Table~\ref{tab:data}. All comments were randomly divided into 10 subsamples, and 9 were used for training and the remaining was used for testing. The cross-validation process is repeated 10 times. For testing, all publication-related entities were identified with the trained models.

For comments with citations, there were 11.1 identified entities on average. However, there were only 1.2 identified entities on average for comments without citations.
With the set of identified entities in comments, the detection performance was measured in terms of precision, recall, and $F_1$.
From all possible combinations with high $F_1$, the combinations sorted by precision are shown in Table~\ref{tab:combination}. For these combinations, we achieved a high $F_1$, that is, at least 0.87.
Since we are attempting to detect publication citations in a large amount of comments, we consider precision as the most important metric. The combination of `author', `title', `year', and `booktitle\_or\_journal' is found to lead to the highest precision of 0.99.

From this result we consider that NER with our annotated comments works well to detect publication citations, and we train the model \texttt{en\_core\_web\_sm} with all 272 annotated comments for large-scale analysis.

\section{Detection Evaluation}
\label{sec:evaluation}

In this section, we present our evaluation of the detection of publication citations in a large amount of source code comments and our manual evaluation on a statistically representative sample of the detected publication citations.

\subsection{Automated Detection}

The trained model prepared in Section~\ref{sec:train} was used to detect publication citations in all distinct 319,438,977 source code comments.
To decrease incorrectly detected comments, the set of four entities (\textbf{author}, \textbf{title}, \textbf{year}, and \textbf{booktitle\_or\_journal}), which shows the highest precision in Table~\ref{tab:combination}, was used as detection criterion.

Different from the small sample used for validation in Section~\ref{sec:train}, with the large amount of data, we found that the criterion with the four entities only was not always sufficient to detect publication citations accurately. Some larger source code comments were incorrectly found to include the four types of entities in separate parts of the comment, even if they did not contain any publication citations.
To mitigate such mis-identification, we measure the distance between detected entities, and remove source code comments with a distance of more than 10 (including spaces, characters, and special characters) for the largest gaps.

\begin{table}
    \centering
    \caption{Detected publication citations and their statistically representative sample}
    \label{tab:overviewofsample}
    \begin{tabular}{lrr}
        \toprule
         & \textbf{\#} & \textbf{\%}  \\
        \midrule
        all comments & 319,438,977 & 100\% \\
        comments satisfying the detection criteria & 11,724 & 0.0037\%   \\
        \midrule
        sample from the above 11,274 comments & 372 & 100\% \\
        cite & 305 & 82\%    \\
        none & 67 & 18\%  \\
        \bottomrule
    \end{tabular}
\end{table}

As shown in Table~\ref{tab:overviewofsample}, we obtained 11,724 distinct comments (C 3,696, C++ 3,371, Java 2,415, JavaScript 197, Python 1,994, PHP 30, and Ruby 21), which is only 0.0037\% of all comments.
Although our detection criteria might miss some comments containing publication citations, we estimate that the percentage of comments with publication citations will not exceed 0.1\%.

\subsection{Sampling and Evaluation}
\label{sec:sample}

Since investigating 11,724 comments manually is not practical, we prepared a statistically representative sample,
so that conclusions would generalize to all 11,724 identified comments.
The sample size was again calculated with a confidence level of 95\% and a confidence interval of 5, to obtain 372 comments.
The second author investigated these 372 comments and found that 305 comments actually included publication citations, which is about 82\% of the sample.

\begin{figure}
\centering
\begin{tikzpicture}
    \begin{axis}[
        title={},
        xlabel={The largest distance},
        xtick={3,4,5,6,7,8,9,10},
        xmin=2, xmax=11,
        ymin=0, ymax=1,
        ymajorgrids=true,
        every axis plot/.append style={line width=1pt},
        legend style={at={(0.97,0.03)},anchor=south east}
    ]
    \addplot[-, color=blue, smooth]
        coordinates{
        (3,1)(4,0.85)(5,0.85)(6,0.83)(7,0.83)(8,0.83)(9,0.83)(10,0.82)
        };
        \addlegendentry{Precision}
    \addplot[-, color=green, smooth]
        coordinates{
        (3,0.11)(4,0.7)(5,0.75)(6,0.86)(7,0.94)(8,0.97)(9,0.98)(10,1)
        };
        \addlegendentry{Recall}
    \addplot[dashed, color=red, smooth]
        coordinates{
        (3,0.21)(4,0.76)(5,0.79)(6,0.84)(7,0.88)(8,0.9)(9,0.9)(10,0.9)
        };
        \addlegendentry{$F_1$}
    \end{axis}
\end{tikzpicture}
\caption{Detection performance with the distance threshold}
\label{fig:threshold}
\end{figure}
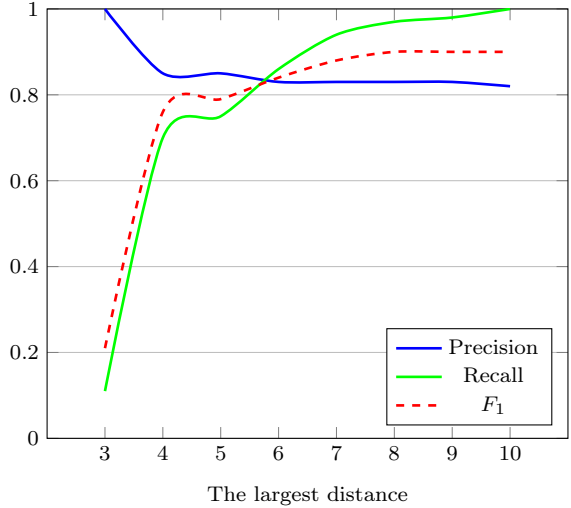

\textbf{Publication detection accuracy.}
We conducted a sensitivity analysis of the distance parameter threshold, that is, precision, recall, and $F_1$ are measured with different thresholds for the largest distance to be included, from 3 to 10.
Figure~\ref{fig:threshold} presents the result of this sensitivity analysis.
Although smaller thresholds make recall worse, bigger thresholds decrease precision gradually.
With the largest distance as 10, we obtained the highest $F_1$ score of 0.90.

\begin{table}
    \centering
    \caption{Identified entities in the 305 sample}
    \label{tab:freqcorrectentity}
    \begin{tabular}{lrrrr}
        \toprule
        & \textbf{correct} & \textbf{(\%)} & \textbf{partially cor.} & \textbf{incor.} \\
        \midrule
        author & 237 & (78\%) & 63 & 5 \\
        title & 233 & (76\%) & 53 & 19 \\
        year & 269 & (88\%) & 36 & 0 \\
        booktitle\_or\_journal & 233 & (76\%) & 38 & 34 \\
        \bottomrule
    \end{tabular}
\end{table}

\textbf{Entity identification accuracy.}
Although we could detect publication citations accurately, it is not clear how well we are able to detect each of the different entities.
The second author again manually validated the four types of entities in the 305 comments that include publication citations.
In a comment, there can be multiple entities for the same type, such as multiple authors for one publication. If all entities of the same type are appropriate, we consider the analysis of this comment to be \textit{correct}. If all entities of the same type are inappropriate, we consider it \textit{incorrect}. We consider it \textit{partially correct} otherwise.

Table~\ref{tab:freqcorrectentity} summarizes the number of comments with the above categories.
As expected, \textbf{year} is the most correctly identified entity, accounting for 88\%. At least one \textit{year} entity was identified correctly in all comments.
The other three entities achieve similar accuracy of \textit{correct}, accounting for 76--78\%.
Considering \textit{partially correct}, part of the \textbf{author} entities were identified accurately in most comments.
Among the four entity types, the entity \textbf{booktitle\_or\_journal} is found to be relatively difficult to identify.

\vspace{2cm}
\begin{figure*}
    \begin{alltt}
TYPE-I Lorentzian, \hlauthor{Becker \textbf{AUTHOR}} , \hlauthor{P. J. \textbf{AUTHOR}} & \hlauthor{Coppens, P. \textbf{AUTHOR}} ( \hlyear{1974 \textbf{YEAR}} ).
\hltitle{Acta Cryst \textbf{TITLE}} . \hlbook{A30 \textbf{BOOKTITLE_OR_JOURNAL}} , \hlvolume{129 \textbf{VOLUME}} ;
    \end{alltt}
    \caption{A successfully detected publication citation that does not include title. The actual title is ``\textit{Extinction within the limit of validity of the Darwin transfer equations. I. General formalism for primary and secondary extinction and their applications to spherical crystals}''}
    \label{fig:missingtitlecomment}
\end{figure*}

\begin{figure*}
    \begin{alltt}
@ file @ ingroup dspSpatLib @ brief TODO @ details TODO @ n @ authors Trond Lossius , 
\hlauthor{Nils Peters \textbf{AUTHOR}} , \hltitle{Timothy Place @ copyright Copyright \textcopyright \textbf{TITLE}}  \hlyear{2011 \textbf{YEAR}} by 
\hlauthor{Trond Lossius \textbf{AUTHOR}} , \hlauthor{Nils Peters \textbf{AUTHOR}} , and Timothy Place @ n This code is 
licensed under the terms of the `` \hlbook{New BSD License \textbf{BOOKTITLE\_OR\_JOURNAL}} '' @ n 
\hlurl{http : creativecommons.org licenses BSD \textbf{URL}} .
     \end{alltt}
\caption{An incorrectly detected publication citation}
\label{fig:exampleofincorrectcitation}
\end{figure*}

Figure~\ref{fig:missingtitlecomment} shows a part of a source code comment with a publication citation correctly detected.
In this citation, the publication title is not present.
Although \textbf{author} and \textbf{year} were identified almost correctly\footnote{``Becker, P. J.'' is one author entity.}, inappropriate entities were identified as \textbf{title} and \textbf{booktitle\_or\_journal}.
However, since the four entity types were included and the largest distance is smaller than 10, this publication citation was correctly detected.
Similar to this example, our approach is able to detect publication citations correctly even in cases where some identified entities are incorrect.

Figure~\ref{fig:exampleofincorrectcitation} shows an example of a comment that was incorrectly detected as containing a publication citation. Similar to this example, parts of comments that include person names and years were prone to be detected as publication-related entities.
Compared to the performance of the detection of publication citations, the results of the identified entities were relatively lower.
Improving our detection criteria and introducing automated validation processes should be promising future directions.

\textbf{Comparison with a benchmark.}
We also performed a comparison between our approach and a pattern-based approach \citep{ambpp.2016.12751}, i.e., the current state of the art in detecting publication references in patents.
This approach uses a group of patterns, i.e., author surname, year of journal, journal volume, and initial page number, to detect publication citations.
In other words, citations are detected only if they contain all of these elements.
The third author manually checked whether publication citations detected by our approach could have been detected by such a pattern for 50 comments in our representative sample.
The result shows that we can outperform the benchmark, since only 54\% (27) of the citations are detected by the baseline approach.
Our approach is able to find a substantial number of referenced publications that the baseline approach would not have been able to detect, partly because the baseline approach does not consider paper titles, journal names, publishers, etc., which our NER approach was able to identify.


\section{Citation Analyses}
\label{sec:analysis}

In this section, we present our findings of an in-depth analysis of the detected publication citations in source code comments.

\begin{table}
    \caption{The number of publication citations in a comment}
    \label{tab:numberofcitationpercomment}
    \centering
    \begin{tabular}{rrr}
        \toprule
        \textbf{\# publications} & \textbf{\# comments} & \textbf{\%}  \\
        \midrule
        1 & 176 & 58\%  \\
        2 & 93 & 30\%   \\
        3 & 21 & 7\%    \\
        4 & 8 & 3\% \\
        5 & 6 & 2\% \\
        6 & 1 & 0\% \\
        \midrule
        \textbf{sum} & \textbf{305} & \textbf{100\%} \\
        \bottomrule
    \end{tabular}
\end{table}

\subsection{Citation Statistics}

To understand the common practices around publication citations in source code comments,
we investigate the number of publications referenced per source code comment.
Table~\ref{tab:numberofcitationpercomment} groups the comments by the number of publications that they cite.
From the sample of 305 comments with publication citations, 58\% referenced only one publication.
Interestingly, more than 40\% of the comments referenced multiple publications.
At most six publications were referenced in one comment.
We conclude that not only single publications but also multiple publications can be made use of as knowledge sources when implementing a piece of source code.

\subsection{Qualitative Analysis}
\label{sec:qa}

To understand the knowledge transfer that happens from academic publications to source code, we conducted a qualitative investigation of the sample of 305 source code comments with publication citations. Based on preliminary investigation, we focused this analysis on the following aspects related to the potential knowledge transfer: publication type (e.g., journal), knowledge type (e.g., algorithm), scope (e.g., source code method), purpose (e.g., reference), knowledge transfer type (e.g., pseudocode $\rightarrow$ source code), and meta information about the corresponding repositories in terms of type and owner.

This qualitative investigation of the identified 305 source code comments with publication citations was conducted by three authors of this work, by assigning labels for each of the above-mentioned aspects to the source code comments. The three annotators first independently annotated a subset of the same 30 source code comments using open coding to design corresponding coding schemas. After consolidating the codes through discussion, we finalised the coding schemas and then calculated kappa agreement among the annotators.
When the kappa agreement was deemed to be good enough ($\kappa > 0.75$), the remaining cases were then coded by a single annotator.
During the annotation, we investigated the comments as well as associated code on GitHub. We found that 9 comments were no longer available on GitHub.
Thus, for our manual coding, we focused the investigation on the 296 available comments.
If there are multiple publications referenced in single comment, we only target the first publications referenced for this qualitative analysis.

\begin{figure}
\centering
\begin{footnotesize}
\begin{tikzpicture}[scale=0.9]
\begin{axis}[
    align=center,
    y=0.5cm,
    x=0.03cm,
    enlarge y limits={abs=0.25cm},
    symbolic y coords={other, book, conference, journal},
    axis line style={opacity=0},
    major tick style={draw=none},
    ytick=data,
    xmin = 0,
    xlabel = \# publication citations,
    nodes near coords,
    nodes near coords align={horizontal},
    point meta=rawx
]
\addplot[xbar,fill=gray,draw=none] coordinates {
    (171,journal)
    (58,conference)
    (43,book)
    (23,other)
};
\end{axis}
\end{tikzpicture}
\end{footnotesize}
\caption{Frequency of referenced academic publications types}
\label{fig:quali-papertype}
\end{figure}
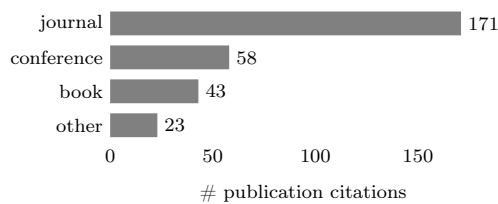

\begin{figure*}
\begin{verbatim}
 * Finds rules according to confirmation measure (Tertius-type algorithm).<br/>
 * <br/>
 * For more information see:<br/>
 * <br/>
 * P. A. Flach, N. Lachiche (1999). Confirmation-Guided Discovery of first-order
 rules with Tertius. Machine Learning. 42:61-95.
 \end{verbatim}
\caption{A code comment referencing the paper \citep{flach2000confirmation-guided} in Machine Learning.}
\label{fig:example-papertype}
\end{figure*}

\subsubsection{Paper Type.}
As our first step in this qualitative analysis, we examined the types of academic publications cited in source code comments.
The same three annotators individually investigated academic publications referenced in the aforementioned 30 source code comments.
In cases where multiple citations are documented, the first one was investigated.
We obtained a kappa agreement of 0.88 or ``almost perfect'' \citep{viera2005understanding}, and the remaining data was investigated by a single rater. Our consolidated coding schema is comprised of the following codes:

\begin{itemize}
    \item \textit{journal}: academic paper published in a journal.
    \item \textit{conference}: academic paper published at a conference.
    \item \textit{book}: scientific book.
    \item \textit{other}: anything that does not fit the previous codes.
    
\end{itemize}

Figure \ref{fig:quali-papertype} shows the results of our qualitative analysis. We observed that most of the publications referenced (58\%) in source code comments are from journals.
For example, as shown in Figure~\ref{fig:example-papertype}, we found a comment in a Java source file\footnote{\url{https://github.com/svn2github/weka/blob/c8366c454e9718d0e1634ddf4a72319dac3ce559/tags/before_classdiscovery_cleanup/weka/associations/Tertius.java#L65}} referencing one article \citep{flach2000confirmation-guided} published in Machine Learning, one of the leading journals for machine learning research.
We also observed that 20\% and 15\% of the referenced publications are conference papers and books, respectively.
Considering the ``other'' code, the results are diverse, e.g., library documentation, reports, and technical memos.

\subsubsection{Knowledge Type.}

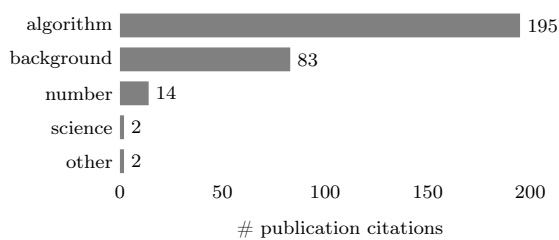
\begin{figure}
\centering
\begin{footnotesize}
\begin{tikzpicture}[scale=0.9]
\begin{axis}[
    align=center,
    y=0.5cm,
    x=0.03cm,
    enlarge y limits={abs=0.25cm},
    symbolic y coords={other,science, number, background, algorithm},
    axis line style={opacity=0},
    major tick style={draw=none},
    ytick=data,
    xmin = 0,
    xlabel = \# publication citations,
    nodes near coords,
    nodes near coords align={horizontal},
    point meta=rawx
]
\addplot[xbar,fill=gray,draw=none] coordinates {
    (195,algorithm)
    (83,background)
    (14,number)
    (2,science)
    (2,other)
};
\end{axis}
\end{tikzpicture}
\end{footnotesize}
\caption{Knowledge types transferred from referenced academic publications.}
\label{fig:quali-knowntype}
\end{figure}

\begin{figure*}
\begin{verbatim}
 * Implementation of the HiCO algorithm, an algorithm for detecting hierarchies
 * of correlation clusters.
 * <p>
 * Reference: E. Achtert, C. Böhm, P. Kröger, A. Zimek:<br />
 * Mining Hierarchies of Correlation Clusters. <br>
 * In: Proc. Int. Conf. on Scientific and Statistical Database Management
 * (SSDBM'06), Vienna, Austria, 2006.
 * </p>
 \end{verbatim}
\caption{A code comment indicating an implementation of the HiCO algorithm \citep{1644305} in a source file.}
\label{fig:example-knowtype}
\end{figure*}

We considered the context of the publication citations, any information in the corresponding publications, and the surrounding source code to investigate characteristics of the knowledge adopted from the academic publications referenced in source code comments.
The kappa agreement between all three annotators was 0.95 (interpreted as ``almost perfect'' \citep{viera2005understanding}).
We use the following five codes to indicate types of knowledge transferred from the referenced academic publications:

\begin{itemize}
    \item \textit{algorithm}: proposed approach, technique, formula, or equation.
    \item \textit{background}: any knowledge from the referenced publication used to express background or additional information related to the source code implementation.
    \item \textit{number}: numbers related to experimental setup (e.g., variables) or numerical findings.
    \item \textit{science}: scientific finding.
    \item \textit{other}: anything that does not fit the previous codes.
\end{itemize}

Figure \ref{fig:quali-knowntype} summarizes the results of this analysis.
The results show that algorithm (66\%) is the most frequent knowledge type transferred from referenced publications.
For example, we found a transfer of the HiCO algorithm proposed in the paper \citep{1644305} to a source file\footnote{\url{https://github.com/elki-project/elki/blob/3e820f9f6382cb82e40955692337737c24f94b54/elki-clustering/src/main/java/de/lmu/ifi/dbs/elki/algorithm/clustering/correlation/HiCO.java#L75}} as shown in Figure \ref{fig:example-knowtype}.
A significant number (28\%) of the referenced academic publications are referred to as background knowledge related to the source code.
The remaining three codes account for only 6\% of the publications in our representative sample.

\begin{figure}
\centering
\begin{footnotesize}
\begin{tikzpicture}[scale=0.9]
\begin{axis}[
    align=center,
    y=0.5cm,
    x=0.045cm,
    enlarge y limits={abs=0.25cm},
    symbolic y coords={no size, file, class, method},
    axis line style={opacity=0},
    major tick style={draw=none},
    ytick=data,
    xmin = 0,
    xlabel = \# source files,
    nodes near coords,
    nodes near coords align={horizontal},
    point meta=rawx
]
\addplot[xbar,fill=gray,draw=none] coordinates {
    (96,method)
    (68,class)
    (49,file)
    (83,no size)
};
\end{axis}
\end{tikzpicture}
\end{footnotesize}
\caption{Extent of transferred knowledge implemented in source files.}
\label{fig:quali-knownsize}
\end{figure}
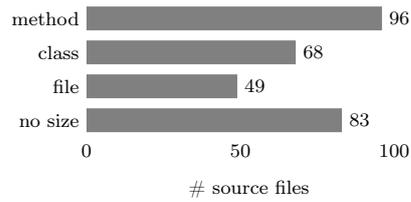
\begin{figure*}
\begin{verbatim}
 Many wavelet coefficient thresholding approaches have been proposed.  By
 default, ``denoise_wavelet`` applies BayesShrink, which is an adaptive
 thresholding method that computes separate thresholds for each wavelet
 sub-band as described in [1]_.
 References
 ----------
 .. [1] Chang, S. Grace, Bin Yu, and Martin Vetterli. "Adaptive wavelet
    thresholding for image denoising and compression." Image Processing,
    IEEE Transactions on 9.9 (2000): 1532-1546.
    DOI: 10.1109/83.862633
 \end{verbatim}
\caption{A code comment with paper citations found within the body of a function in a source file.}
\label{fig:example-scope}
\end{figure*}

\subsubsection{Scope.}
The following list shows the four codes that emerged from our analysis of the scope of the knowledge transferred from academic publications to source code, along with a short description. The kappa agreement was 0.88 or ``almost perfect'' \citep{viera2005understanding}.

\begin{itemize}
    \item \textit{method}: the transferred knowledge is implemented in a method or function in the source file. 
    \item \textit{class}: the transferred knowledge is implemented in a class in the source file. 
    \item \textit{file}: the transferred knowledge is implemented throughout an entire file. 
    \item \textit{no size}: the transferred knowledge is not implemented in any part of the source file, or no knowledge is transferred from the referenced publication. 

\end{itemize}

Figure \ref{fig:quali-knownsize} shows the scope of the knowledge transfer.
We found that a large number (32\%) of the source files that reference academic publications implement the transferred knowledge in a method or function.
An example is shown in Figure \ref{fig:example-scope}. 
In this case, developers explicitly implemented a method described in the paper \citep{862633} in the \texttt{denoise\_wavelet} function\footnote{\url{https://github.com/scikit-image/scikit-image/blob/51f598aaedc73ef180913c670d2c20a8032aaf1e/skimage/restoration/_denoise.py#L497}}.
Also, 23\% of the source files implement the knowledge in a class and 17\% of the source files implement the knowledge throughout an entire file.
Interestingly, 28\% of the source files do not implement knowledge from the publications in any part of the source file. Instead, they reference the publications only for indicating an implementation's background knowledge.

\begin{figure}
\centering
\begin{footnotesize}
\begin{tikzpicture}[scale=0.9]
\begin{axis}[
    align=center,
    y=0.5cm,
    x=0.025cm,
    enlarge y limits={abs=0.25cm},
    symbolic y coords={related, reference},
    axis line style={opacity=0},
    major tick style={draw=none},
    ytick=data,
    xmin = 0,
    xlabel = \# publication citations,
    nodes near coords,
    nodes near coords align={horizontal},
    point meta=rawx
]
\addplot[xbar,fill=gray,draw=none] coordinates {
    (213,reference)
    (83,related)
};
\end{axis}
\end{tikzpicture}
\end{footnotesize}
\caption{Frequency of citation purposes.}
\label{fig:quali-purpose}
\end{figure}
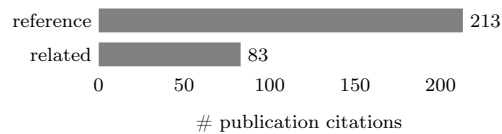

\begin{figure*}
\begin{verbatim}
 * In this implementation, step c is not performed using the usual Chien search.
 * Instead, an alternative approach described in [1] is used. It consists in
 * factoring the error locator polynomial using the Berlekamp Trace algorithm
 
 * [1] B. Biswas, V. Herbert. Efficient root finding of polynomials over fields
 * of characteristic 2, in: Western European Workshop on Research in Cryptology
 * - WEWoRC 2009, Graz, Austria, LNCS, Springer, July 2009, to appear.
 \end{verbatim}
\caption{A code comment referencing the knowledge source \citep{Biswas2009}.}
\label{fig:example-purpose}
\end{figure*}

\subsubsection{Purpose.}
The following list shows the two codes that emerged from our analysis of the purpose of referencing academic publications in source code comments, along with a short description. 
We obtained a kappa agreement of 0.93, interpreted as ``almost perfect'' \citep{viera2005understanding}.

\begin{itemize}
    \item \textit{reference}: the source code comment explicitly indicates that the referenced publication is the official source of some aspect of the source code (e.g., algorithm).
    \item \textit{related}: the referenced publication adds related or additional information to the source code.

\end{itemize}

Figure \ref{fig:quali-purpose} shows the results of our qualitative analysis of the 296 publication citations in representative source code comments.
The results show that the main purpose of providing publication citations is to reference the knowledge source that was used during the implementation, covering more than three quarters of the citations in our sample.
As shown in Figure \ref{fig:example-purpose}, the referenced academic publication \citep{Biswas2009} is considered as the source of the approach implemented in the source file\footnote{\url{https://github.com/Tomoms/helium_kernel/blob/a20b5d0758d5308803394385a8eb753b6b24d345/lib/bch.c#L1}}.
The remaining 28\% of these citations are aimed at providing additional knowledge related to the source code.

\subsubsection{Knowledge Transfer.}
We use the following eight codes to indicate types of knowledge transfer from the referenced academic publications to software development, i.e., source files.
The kappa agreement was 0.78 or ``substanial'' agreement \citep{viera2005understanding}.

\begin{itemize}
    \item \textit{formulas $\rightarrow$ source code}: formulas or equations presented in the referenced publication are transformed into source code. 
    \item \textit{pseudocode $\rightarrow$ source code}: pseudocode presented in the referenced publication is transformed into source code. 
    \item \textit{description $\rightarrow$ source code}: textual descriptions presented in the referenced publication are transformed into source code. 
    \item \textit{numbers $\rightarrow$ hardcoded values}: numbers related to experimental setup (e.g., variables) or numerical findings reported in the referenced publication are added to the source code as hardcoded values.
    \item \textit{scientific finding $\rightarrow$ hardcoded rule}: scientific findings presented in the referenced publication are implemented in the source code as a hardcoded rule.
    \item \textit{documentation}: knowledge from the referenced publication is transferred for documentation only.
    \item \textit{paper not available}: we cannot determine the type of knowledge transfer since the referenced publication is not open-access.
    \item \textit{no transfer}: no knowledge from the referenced publication is transferred to the source code.
\end{itemize}

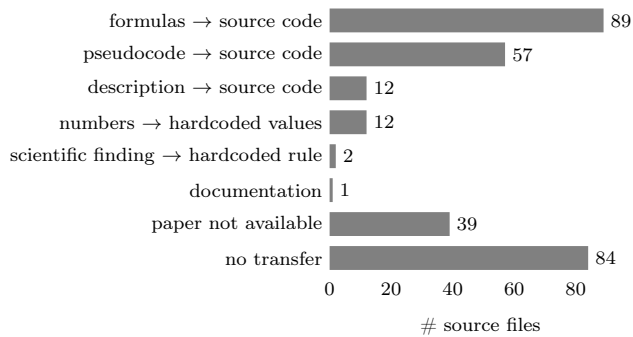
\begin{figure}
\centering
\begin{footnotesize}
\begin{tikzpicture}[scale=0.9]
\begin{axis}[
    align=center,
    y=0.5cm,
    x=0.045cm,
    enlarge y limits={abs=0.25cm},
    symbolic y coords={no transfer,paper not available,documentation,scientific finding $\rightarrow$ hardcoded rule,numbers $\rightarrow$ hardcoded values,description $\rightarrow$ source code,pseudocode $\rightarrow$ source code, formulas $\rightarrow$ source code},
    axis line style={opacity=0},
    major tick style={draw=none},
    ytick=data,
    xmin = 0,
    xlabel = \# source files,
    nodes near coords,
    nodes near coords align={horizontal},
    point meta=rawx
]
\addplot[xbar,fill=gray,draw=none] coordinates {
    (89,formulas $\rightarrow$ source code)
    (57,pseudocode $\rightarrow$ source code)
    (12,description $\rightarrow$ source code)
    (12,numbers $\rightarrow$ hardcoded values)
    (2,scientific finding $\rightarrow$ hardcoded rule)
    (1,documentation)
    (39,paper not available)
    (84,no transfer)
};
\end{axis}
\end{tikzpicture}
\end{footnotesize}
\caption{Knowledge transfer from referenced publications to source code.}
\label{fig:quali-transfer}
\end{figure}

\begin{figure*}
\begin{verbatim}
 .. [1] Max Jaderberg, Karen Simonyan, Andrew Zisserman,
        Koray Kavukcuoglu (2015):
        Spatial Transformer Networks. NIPS 2015,
        http://papers.nips.cc/paper/5854-spatial-transformer-networks.pdf
        
  Here we set up the layer to initially do the identity transform, similarly
  to [1]_. Note that you will want to use a localization with linear output.
  If the output from the localization networks is [t1, t2, t3, t4, t5, t6]
  then t1 and t5 determines zoom, t2 and t4 determines skewness, and t3 and
  t6 move the center position.
 \end{verbatim}
\caption{A code comment indicating a conversion of spatial transformer networks formulas \citep{NIPS2015_5854} into source code.}
\label{fig:example-transfer}
\end{figure*}

As shown in Figure \ref{fig:quali-transfer}, we observed that formulas or equations from the referenced academic publications are frequently implemented in the source code, accounting for 30\% of the cases in our sample, followed by pseudocode implementation (19\%).
For example, we found an implementation of spatial transformer networks formulas \citep{NIPS2015_5854} in a source file\footnote{\url{https://github.com/Lasagne/Lasagne/blob/7992faa80fa5233a786e2582a605e854cea7d1cf/lasagne/layers/special.py#L355}} as shown in Figure \ref{fig:example-transfer}.
Some source files (28\%) with publication citations have no outside knowledge transferred to their source code since the corresponding publications provide only background or additional information associated with them.
Since a significant number (13\%) of the publications referenced in the source files are hidden behind paywalls, especially books and academic papers published in journals, we cannot investigate knowledge transfer in these cases.

\subsubsection{Repository Type.}

We also investigated the domains of the GitHub repositories that reference academic publications with qualitative analysis.
The three annotators first individually checked the same 30 cases, similar to previous sections, and discussed the free-form codes to derive new codes.
Then, the same 30 cases are independently re-coded by all three annotators, using the new coding guide.
Since we obtained kappa agreement of 0.86 (interpreted as ``almost perfect'' \citep{viera2005understanding}), a single annotator investigated the remaining cases.
We present our final coding guide below.

\begin{itemize}
    \item \textit{machine learning}: the repository is focused on machine learning.
    \item \textit{network/os}: the repository is focused on operating systems and/or network structures, including their applications. 
    \item \textit{computer vision}: the repository implements algorithms for computer vision tasks, involving images or videos.
    \item \textit{simulation}: the repository is related to simulation technologies.
    \item \textit{science}: the repository is associated with scientific applications.
    \item \textit{data science}: the repository is concerned with analyzing data to gain knowledge or insights (e.g., data mining).
    \item \textit{other}: anything that does not fit the previous codes.
\end{itemize}

\begin{figure}
\centering
\begin{footnotesize}
\begin{tikzpicture}[scale=0.9]
\begin{axis}[
    align=center,
    y=0.5cm,
    x=0.055cm,
    enlarge y limits={abs=0.25cm},
    symbolic y coords={other,data science,simulation,computer vision,network/os,machine learning,science},
    axis line style={opacity=0},
    major tick style={draw=none},
    ytick=data,
    xmin = 0,
    xlabel = \# GitHub repositories,
    nodes near coords,
    nodes near coords align={horizontal},
    point meta=rawx
]
\addplot[xbar,fill=gray,draw=none] coordinates {
    (61,science)
    (53,machine learning)
    (35,network/os)
    (26,computer vision)
    (25,simulation)
    (12,data science)
    (84,other)
};
\end{axis}
\end{tikzpicture}
\end{footnotesize}
\caption{Frequency of software domains in our sample.}
\label{fig:quali-repotype}
\end{figure}
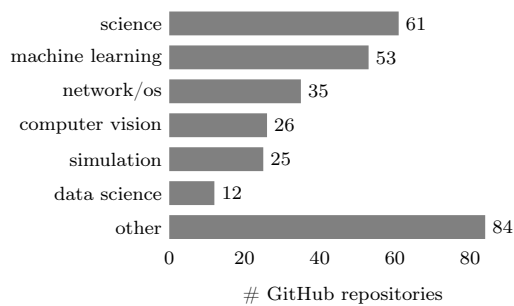

Figure \ref{fig:quali-repotype} shows the results of our qualitative analysis.
We found that the GitHub repositories that reference academic publications are frequently focused on scientific applications, accounting for 21\% of the repositories in our sample, followed by machine learning (18\%) and operating system and/or network structures (12\%).
This repository\footnote{\url{https://github.com/BALL-Project/ball}} is an example of a repository associated with scientific applications, i.e., biochemical algorithms, and containing links to academic publications.
Furthermore, we observed that 84 (28\%) of these repositories belong to the ``other'' code, indicating the diversity of GitHub repositories with publication citations.

\subsubsection{Repository owner.}

To understand the relationship between GitHub repositories and the academic papers they cite, we investigated whether the repository owner is one of authors of the referenced academic publications by comparing the names shown on their GitHub profile page to the author names of the corresponding publications.
If the repository owners are organizations or companies, we investigated their members.
The kappa agreement was 0.97 or ``almost perfect'' \citep{viera2005understanding}.
We show our coding guide below, along with a short description.

\begin{itemize}
    \item \textit{identical}: the repository owner is one of the authors of the referenced publication.
   \item \textit{not identical}: the repository owner is not one of the authors of the referenced publication.
\end{itemize}

\begin{figure}
\centering
\begin{footnotesize}
\begin{tikzpicture}[scale=0.9]
\begin{axis}[
    align=center,
    y=0.5cm,
    x=0.02cm,
    enlarge y limits={abs=0.25cm},
    symbolic y coords={identical, not identical},
    axis line style={opacity=0},
    major tick style={draw=none},
    ytick=data,
    xmin = 0,
    xlabel = \# GitHub repositories,
    nodes near coords,
    nodes near coords align={horizontal},
    point meta=rawx
]
\addplot[xbar,fill=gray,draw=none] coordinates {
    (281,not identical)
    (15,identical)
 
};
\end{axis}
\end{tikzpicture}
\end{footnotesize}
\caption{Relationship between GitHub repository owners and publication authors.}
\label{fig:quali-author}
\end{figure}
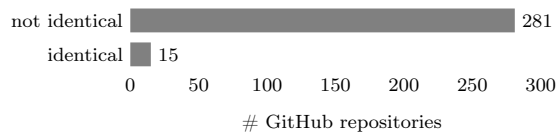

\begin{figure*}
\begin{verbatim}
 Compute the window-convoled power spectrum multipoles, for a data set
 with non-trivial survey geometry.
 References
----------
 * Bianchi, Davide et al., `Measuring line-of-sight-dependent Fourier-space
 clustering using FFTs`, MNRAS, 2015
 * Scoccimarro, Roman, `Fast estimators for redshift-space clustering`, Phys.
 Review D, 2015
 \end{verbatim}
\caption{A code comment referencing publications whose authors are different from the repository owner.}
\label{fig:example-owner}
\end{figure*}

Figure \ref{fig:quali-author} shows the results of our qualitative analysis of the 296 GitHub repositories. 
The results show that for the vast majority of the publication citations in GitHub repositories, the owner of the repository is not associated with the publication as an author.
Figure \ref{fig:example-owner} shows an example of an academic publications referenced by a repository owner \footnote{\url{https://github.com/bccp/nbodykit/blob/c90c009c9b7ab69447a7199a57605837dad9ae91/nbodykit/algorithms/convpower.py#L404}} who is different from the publication authors.

\subsection{Popular Publications}

\begin{table*}
    \centering
    \caption{Top 20 referenced publications}
    \label{tab:top20papers}
    \resizebox{\columnwidth}{!}{%
    \begin{tabular}{rp{21pc}rp{7pc}r}
        \toprule
        \textbf{rank} & \textbf{publication} & \multicolumn{2}{l}{\textbf{\# code citations}} & \textbf{\# paper citations} \\
        \midrule
        1 & C. Loeffler, A. Ligtenberg, G.S. Moschytz, \textbf{Practical fast 1-D DCT algorithms with 11 multiplications},  Proceedings of 1989 International Conference on Acoustics, Speech, and Signal Processing (1989) & 146 & (C 105, C++ 19, Java 1, JavaScript 21) & 1,049 \\
        1 & S. Reiter, A. Vogel, I. Heppner, M. Rupp, G. Wittum, \textbf{A massively parallel geometric multigrid solver on hierarchically distributed grids}, Computing and Visualization in Science (2013) & 146 & (C++ 146) & 49 \\
        3 & S. Tomov, J. Dongarra, \textbf{Accelerating the reduction to upper Hessenberg form through hybrid GPU-based computing}, University of Tennessee Computer Science Technical Report (2009) & 75 & (C++ 75) & 31  \\
        4 & V.I. Lebedev, D.N. Laikov, \textbf{A quadrature formula for the sphere of the 131st algebraic order of accuracy}, Doklady Mathematics (1999) & 71 & (C++ 71) & 522  \\
        5 & R. Sedgewick, \textbf{Algorithms, 2nd Edition}, Addison-Wesley (1988) & 67 & (C 61, C++ 6) & 5,335 \\
        5 & E.R. Fiala, D.H. Greene, \textbf{Data compression with finite windows}, Communications of the ACM (1989) & 67 & (C 61, C++ 6) & 300 \\
        7 & P. L'Ecuyer, \textbf{Maximally equidistributed combined Tausworthe generators}, Mathematics of Computation (1986) & 63 & (C 59, C++ 4) & 326 \\ 
        8 & M. Matsumoto, T. Nishimura, \textbf{Mersenne twister: a 623-dimensionally equidistributed uniform pseudo-random number generator}, ACM Transactions on Modeling and Computer Simulation (1998) & 57 & (C 24, C++ 14, Java 19) & 6,752 \\
        8 & P. L'Ecuyer, \textbf{Tables of maximally equidistributed combined LFSR generators}, Mathematics of Computation (1999) & 57 & (C 56, C++ 1) & 225  \\
        10 & J.P. Snyder, \textbf{Map projections--A working manual}, U.S. Geological Survey Professional Paper (1987) & 55 & (C++ 2, JavaScript 53) & 1,736 \\
        10 & P. Deutsch, \textbf{DEFLATE compressed data format specification version 1.3}, RFC 1951 (1996) & 55 & (C 51, C++ 4) & 714 \\
        12 & J.M. Robson, \textbf{Bounds for some functions concerning dynamic storage allocation}, Journal of the ACM (1974) & 51 & (C 44, C++ 7) & 84 \\
        13 & J.C.R. Bennett, H. Zhang, \textbf{Hierarchical packet fair queueing algorithms}, IEEE/ACM Transactions on Networking (1997) & 50 & (C 50) & 605 \\
        13 & I. Stoica, H. Abdel-Wahab, \textbf{Earliest eligible virtual deadline first: A flexible and accurate mechanism for proportional share resource allocation}, Technical Report (1995) & 50 & (C 50) & 60 \\
        15 & M. Matsumoto, Y. Kurita, \textbf{Twisted GFSR generators II}, ACM Transactions on Modeling and Computer Simulation (1994) & 49 & (C 47, C++ 2) & 230 \\
        16 & M. Matsumoto, Y. Kurita, \textbf{Twisted GFSR generators}, ACM Transactions on Modeling and Computer Simulation (1992) & 47 & (C 47) & 218 \\
        17 & S. Muchnick, \textbf{Advanced compiler design and implementation}, Academic Press, Morgan Kaufmann Publishers (1997) & 43 & (C 43) & 3,908 \\
        17 & R.J. Gowersk, M. Linke, J. Barnoud, T.J.E. Reddy, M.N. Melo, S.L. Seyler, J. Domanski, D.L. Dotson, S. Buchoux, I.M. Kenney, O. Beckstein, \textbf{MDAnalysis: a Python package for the rapid analysis of molecular dynamics simulations}, Proceeding of 15th Python in Science Conference (2016) & 43 & (Python 43) & 167   \\
        17 & Y. Wang, I.H. Witten, \textbf{Modeling for optimal probability prediction}, Proceedings of 19th International Conference on Machine Learning (2002) & 43 & (Java 43) & 60  \\
        17 & Y. Wang, \textbf{A new approach to fitting linear models in high dimensional spaces}, PhD Thesis, University of Waikato, NZ (2000) & 43 & (Java 43) & 48 \\
        \bottomrule
    \end{tabular}%
    }
\end{table*}

\begin{table}
    \centering
    \caption{Frequently referenced journals or conferences}
    \label{tab:topjournals}
    \begin{tabular}{rlr}
        \toprule
        \textbf{rank} & \textbf{journal or conference} & \textbf{\# papers}  \\
        \midrule
        1 & ACM T MATH SOFTWARE & 7 \\
        2 & COMMUN ACM & 4    \\
        3 & ACM T MODEL COMP SIM & 3  \\
         & IEEE ACM T NETWORK & 3 \\
         & INT CONF MACHINE LEARNING & 3    \\
         & MATH COMP & 3  \\
        7 & ACM SIGCOMM & 2 \\
         & ADDISON-WESLEY & 2  \\
         & CONF UNC ARTIFICIAL INTELLIGENCE & 2   \\
         & EURO CONF MACHINE LEARNING & 2   \\
         & IEEE INFOCOM & 2    \\
        \bottomrule
    \end{tabular}
\end{table}

Across all source code comments in our entire dataset, which publications have been referenced most frequently? To answer this question, we collected all entities identified as \textbf{title} from the 11,274 comments that satisfy the detection criteria.
Those entities were sorted by the number of comments that include them. Entities that appear at least 20 times were validated manually, and 93 titles were obtained.

Table~\ref{tab:top20papers} shows the top 20 most referenced publications cited in source code comments.
We found various types of publications in the top 20, that is, 10 journal articles, 3 conference papers, 3 books, 2 technical reports, 1 Request for Comments (RFC), and 1 PhD thesis.
This result demonstrates the effectiveness of our NER approach, as our preliminary manual investigation with predetermined keywords (Section~\ref{sec:manual}) could not detect various publications types.
Table~\ref{tab:top20papers} also presents the number of citations in academia, based on \texttt{Google Scholar} accessed in April 2020. 
Although some publications have lower impact in academia, they can have higher impact in software development.

Table~\ref{tab:topjournals} summarizes the number of distinct referenced publications from the 93 most popular publications in terms of their journals or conferences.
\textsf{ACM TRANSACTIONS on MATHEMATICAL SOFTWARE} and \textsf{COMMUNICATION of the ACM} were found to be the most referenced journals, which is similar to Table~\ref{tab:journal}.

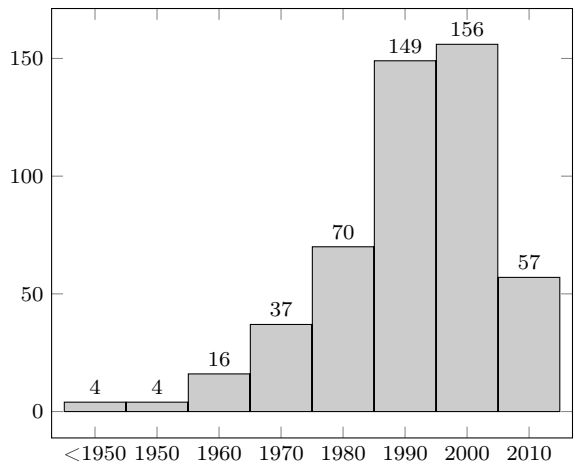
\begin{figure}
    \centering
    \begin{tikzpicture}
        \begin{axis}[
            compat=newest,
            symbolic x coords={\textless1950,1950,1960,1970,1980,1990,2000,2010},
            bar width=23pt,
            nodes near coords,
            every node near coord/.append style={font=\footnotesize},
            xtick=data,
            ticklabel style={font=\small},
          ]
            \addplot[ybar,fill=black!20] 
            coordinates {
                (\textless1950,4)
                (1950,4)
                (1960,16)
                (1970,37)
                (1980,70)
                (1990,149)
                (2000,156)
                (2010,57)
            };
        \end{axis}
    \end{tikzpicture}
    \caption{Frequency of citations per decade from sample}
    \label{fig:freqpaperdecadesample}
\end{figure}

\subsection{Publication Years}

To understand the time aspect of academic knowledge sources used for software development, we analyzed publication years in the sample of 305 publication citations. The second author manually confirmed all publication years. As seen in Table~\ref{tab:numberofcitationpercomment}, there can be multiple publication years in a single comment.
Figure~\ref{fig:freqpaperdecadesample} shows the histogram of publication years.
Most publications referenced in source code comments have been published in the 1990s and 2000s.

\section{Discussion}

In addition to gaining insights into the kinds of knowledge transfer that happen from academic publications to source code, our analysis of publication citations in source code comments also revealed anecdotal issues worthy of further investigation. From the top 20 referenced publications shown in Table~\ref{tab:top20papers}, we observed three related papers, that is, \textbf{Twisted GFSR generators} by Matsumoto and Kurita (1992 and 1994)~\citep{Matsumoto:1992:TGG:146382.146383,Matsumoto:1994:TGG:189443.189445} and \textbf{Mersenne twister} by Matsumoto and Nishimura (1998)~\citep{Matsumoto:1998:MTE:272991.272995}.
The \texttt{Mersenne twister} is extended from the previous \texttt{Twisted GFSR}~\citep{Matsumoto:1998:MTE:272991.272995}, and is the industry standard algorithm for generating random samples~\citep{Marsland:2014:MLA:2692349}.
Programs of \texttt{Mersenne twister} implemented by the authors are provided\footnote{\url{http://www.math.sci.hiroshima-u.ac.jp/~m-mat/MT/emt.html}}.

Considering \texttt{Mersenne twister} is improved from \texttt{Twisted GFSR} and is the industry standard, referencing the paper of \texttt{Mersenne twister} to implement the algorithm seems to be appropriate, instead of (only) referencing the older papers of \texttt{Twisted GFSR}. For the number of citations in academic papers, only the paper of \texttt{Mersenne twister} has obtained a large amount of citations (more than 6,200). However, for citations in code comments, the differences in the number of citations among the three papers are surprisingly small.

We searched for the paper title ``Twisted GFSR generators'' in all comments including duplicates from all 2,482 C repositories (Table~\ref{tab:comments}). In total 142 comments were obtained. All these comments have more or less the same content, which is likely the result of code reuse. Because of several modifications for different typos, variations in the text exist, resulting in 47 distinct comments\footnote{This is a minor documentation issue.}.
All of these comments reference both of the \texttt{Twisted GFSR} papers~\citep{Matsumoto:1992:TGG:146382.146383,Matsumoto:1994:TGG:189443.189445} but they do not reference the \texttt{Mersenne twister} paper~\citep{Matsumoto:1998:MTE:272991.272995}, although their associated code is for random number generation.
It appears that the original developer implemented code based on \texttt{Twisted GFSR} and the code has been reused by many active repositories. 
If developers are not intentionally avoiding the \texttt{Mersenne twister} algorithm, there seems to be an issue of potentially obsolete knowledge. 
If these repositories update to \texttt{Mersenne twister}, there could be practical impact for developers and users.

In general, our findings can be summarized into recommendations for software developers and software engineering researchers.

Our recommendation for software developers is:
\begin{itemize}
    \item \textit{Be aware of state-of-the-art academic achievements} to maintain and improve source code. Past publications might turn out to be buggy, or new algorithms can overcome shortcomings of older algorithms.
\end{itemize}

We also consider implications for software engineering researchers:
\begin{itemize}
    \item \textit{Further studies of clues regarding the intentions of developers in source code comments}.
    As previous studies revealed, source code comments contain different types of clues to developers' intentions, such as personal and team tasks~\citep{Storey:2008:TBE:1368088.1368123}, technical debt~\citep{potdar2014exploratory}, and external information sources via links~\citep{Hata2019}. Publication citations can be another clue to developers' knowledge sources. Deepening the understanding of such clues or studying different types of clues is required to further understand developers' intentions and their code.
    \item \textit{Tool support for knowledge transfer from academia to software development}.
    Although paying attention to recent academic achievements can be a good custom for developers, it is not practical to cover various research topics related to large amounts of code resources. Tools or systems to support knowledge transfer and maintain knowledge-code traceability chains could be practically useful.
\end{itemize}

\section{Threats to Validity}
Several threats to the \textit{construct validity} exist in our study.
Since our approach for citation detection requires four types of entities, publication citations that lack those types could not be detected and were ignored in our citation analysis. For example, previous work on links in source code comments reported that there are references to academic publication only with links~\citep{Hata2019}. Such publication citations cannot be detected with Named Entity Recognition.

Threats to the \textit{external validity} exist in our data preparation.
Since our dataset comes only from GitHub, we cannot generalize our findings to industry or free/libre and open source software in general. Moreover, although we targeted seven common programming languages, other programming languages might have different characteristics.

To minimize threats to \textit{reliability}, we make our dataset publicly available.
We provide our data in an online appendix including our source code files and identified publication references (see Section~\ref{ssec:appendix}).

\section{Related Work}

Since the scientific citation index was proposed by \cite{Garfield649}, the study of academic paper citations and citation analysis have become important fields on their own.
Paper citation has been studied from multiple perspectives, such as categorizing citation profiles \citep{Chakraborty:2015:CSC:2817191.2701412} and analyzing how scholars cite papers \citep{Bornmann:10.1108/00220410810844150, Burrell:10.1002/asi.10207}.

\cite{Chakraborty:2015:CSC:2817191.2701412} present an analysis that characterizes the important categories of scientific citations in computer science.
The authors built a model for classifying papers into temporary and perennial, so that the behavior of numerous citation profiles can be replicated.

An investigation of citing behavior conducted by \cite{Bornmann:10.1108/00220410810844150} reveals that scientists' motivation to cite papers is not only for intellectual acknowledgement, but also other non-scientific factors, which is a relevant indicator for the growth of citation numbers.
A study on citation behavior prediction \citep{Burrell:10.1002/asi.10207} has shown that the number of citations in the future will increase linearly compared to the current number.
\cite{ambpp.2016.12751} investigated the impact of academic institutions' location on the citation behavior of firms.
Their findings suggest that academic research conducted in commercial R\&D hubs in the same field are more likely to be referenced by firms than the same research conducted outside such a hub.

Bibliometric topics have also been investigated in several studies, for instance analyzing the reason of citing papers \citep{Moravcsik10.2307/284557,Small10.2307/284908,Teufel:2009:ASC:1654595.1654612}, evaluating the impact of academic papers \citep{Shi:2010:CHI:1816123.1816131} and analyzing the social network of citations \citep{Dawson:2014:CSF:2567574.2567585}.

In addition, several investigations on referencing artefacts other than research papers from software systems have been conducted.
\cite{Gomez2013} investigated link sharing on Stack Overflow to understand how software developers discover and disseminate innovations.
\cite{Rath2018} investigated the appearance of links to issue tracking systems in commit comments.
Their findings indicate that commit comments are less likely to contain links to specific issues.
Inspired by this problem, they proposed an approach to automatically recover missing links between commits and issues by searching issues associated with a given commit.
\cite{Alqahtani2017} proposed an approach to automatically link online resources from dependent components in a system, to investigate the impact of security vulnerabilities within and across system boundaries.
\cite{Chen2015} also proposed a tool using similarity of code fragments to link between source code issues and relevant Stack Overflow questions which provide solutions and explanations of the problematic source code.
To recover links between software artifacts whose descriptions are different in terms of language or abstraction level, \cite{RyosukeTSUCHIYA20192018EDP7331} proposed a method based on an assumption that the links between two software artifacts can be recovered transitively by tracking links among multiple software artifacts.
They found that the method would be more effective compared to those methods using textual similarity when language type and vocabulary of software artifacts are different.  

Recently, traceability links between source code and documentation have been investigated.
\cite{Scanniello2018} conducted several controlled experiments of program comprehension tasks with UML models produced in a requirements engineering phase and a
design phase to analyze the contribution made by the UML models to the comprehensibility of source code.
They found that models produced in the design phase could improve source code comprehensibility of software developers. 
\cite{Antoniol2000} proposed an approach to identify links between design documents and source files by comparing attribute names of a class to its original class definition in
design documents.
The approach could help developers to deal with inconsistencies and support the evolution of these links.
Similarly, \cite{Rahimi2019} focused on the evolution of links between source files and requirements documents.
They presented an approach to automatically update these links based on rules corresponding to change scenarios from semantic differences of source code.
They revealed that the approach would be effective to evolve trace links.
The work most relevant to ours from the software engineering research community is the one conducted by \cite{Hata2019} in which reference links in source code comments were investigated.
Their empirical analysis on more than 20,000 Git repositories revealed that links to academic publications was one of 19 different kinds of links that emerged from their analysis.
Their analysis of link targets referenced in
source code comments also revealed that the links to academic publications typically belong to the sometimes or rarely linked domains.
They also found that some links to academic publications are dead or inaccessible indicating problems with maintaining these links.

In this work, we zoom in on the knowledge transfer through the links between software development and academic publications.
Furthermore, we propose an approach for the automated detection of publication citations in source code comments, and take a deeper look at the publication citations to understand common practices around publication citations in source code comments, characteristics of popular publications, and time aspects of academic knowledge sources referenced in source code comments.

\section{Conclusion}
To understand the contribution of academic publications to software development, we
(i) trained a Named Entity Recognition model to automatically detect publication citations in source code comments;
(ii) conducted a large-scale study with 319,438,977 distinct comments extracted from 25,925 active repositories written in seven languages; and
(iii) performed a qualitative study of detected publication citations to understand knowledge transfer from academia to software development.

Our study has shown that software development cannot be separated from the achievements in academia.
Based on this work, there are many open avenues for future work, such as further studies of clues to the intentions of developers in source code comments and tool support for knowledge transfer and traceability from academia to software development.


%
%


%
%


\end{sloppy}
\end{document}